%% file: main.tex
\documentclass[aps,prl,reprint,superscriptaddress,floatfix]{revtex4-2}

\usepackage{amsmath}
\usepackage{amssymb}
\usepackage{bm}
\usepackage{graphicx}
\usepackage{slashed}
\usepackage{etoolbox}
\usepackage{placeins}
\usepackage[colorlinks=true,allcolors=blue]{hyperref}

\begin{document}

\title{Next-to-leading order QCD corrections to fully charm tetraquark radiative decays}

\author{Xiang Li}
\email{lix-PHY@pku.edu.cn}
\affiliation{School of Physics, Peking University, Beijing 100871, China}

\author{Yan-Qing Ma}
\email{yqma@pku.edu.cn}
\affiliation{School of Physics, Peking University, Beijing 100871, China}
\affiliation{Center for High Energy Physics, Peking University, Beijing 100871, China}

\author{Bo-Xuan Shi}
\email{shibx@pku.edu.cn}
\affiliation{Center for High Energy Physics, Peking University, Beijing 100871, China}

\author{Sheng-Qi Zhang}
\email{shqzhang@pku.edu.cn}
\affiliation{Center for High Energy Physics, Peking University, Beijing 100871, China}

\begin{abstract}
We present the first complete next-to-leading-order (NLO) QCD calculation of the fully charm tetraquark radiative decays $T_{4c}(0^{++},2^{++})\to\gamma J/\psi$ and $T_{4c}(1^{+-})\to\gamma\eta_c$ at leading order in the relative velocity. The NLO widths are $2.3$--$3.1$ times the leading-order (LO) widths for the long-distance matrix element (LDME) inputs considered here, with $m_c=1.5$ GeV and $\mu_R=2m_c$. In the scalar channel, the NLO corrections change the relative contributions of the two color-spin components and their interference. For the tensor channel, the longitudinal fraction $f_L$ of the daughter $J/\psi$ changes from $0.11$ at LO to $0.18$ at strictly expanded NLO, with sensitivity to higher-order corrections. These NLO results provide valuable theoretical input for searches for states predicted to lie near their respective lowest dicharmonium thresholds and for probing the spin and short-distance structure of tetraquark candidates.
\end{abstract}

\maketitle

\textit{Introduction}. Fully charm tetraquarks provide a rare opportunity to study how QCD binds four quarks within a nonrelativistic framework. Their heavy constituents permit an expansion in the relative velocity, while the absence of valence light quarks limits the role of light-meson exchange \cite{Anwar:2017toa,Feng:2023agq,Wang:2021kfv}. These features make their spectrum a test of short-distance color interactions and nonperturbative four-quark binding.

The LHCb Collaboration first observed a pronounced structure near $6.9$ GeV in the di-$J/\psi$ spectrum \cite{LHCb:2020bwg}. ATLAS and CMS later confirmed the $X(6900)$ and reported additional structures near $6.6$ and $7.1$ GeV \cite{ATLAS:2023bft,CMS:2023owd}. An NRQCD study of hadroproduction had previously suggested spin 2 for the observed di-$J/\psi$ resonances \cite{Zhang:2020hoh}. Recent CMS analyses favor a common $J^{PC}=2^{++}$ assignment for the $X(6600)$, $X(6900)$, and $X(7100)$ \cite{CMS:2025fpt,CMS:2026tiu}. This assignment supports their interpretation as radial excitations \cite{Zhu:2020xni,Lin:2024olg}, but leaves their microscopic origin unsettled. Compact tetraquarks, hadronic molecules, and threshold effects can produce similar spectra \cite{Cao:2020gul,Bedolla:2019zwg,Wan:2020fsk,Wang:2020wrp,Dong:2020nwy,Gong:2020bmg,Albuquerque:2020hio,Guo:2020pvt}. Discriminating among these possibilities requires complementary decay observables.

For the observed structures above the di-$J/\psi$ threshold, radiative decays into a photon and charmonium offer complementary information on their nature~\cite{Guo:2026ngz,Fu:2025lfo}. The radiative process proceeds through short-distance $c\bar c$ annihilation and probes the quark contact density \cite{Voloshin:2006hh}, while the di-$J/\psi$ spectrum can also be affected by threshold effects and rescattering \cite{Guo:2020pvt}. In NRQCD, the contact densities appear as long-distance matrix elements (LDMEs) of definite color-spin configurations, factorized from perturbative short-distance coefficients \cite{Bodwin:1994jh}. Together with the di-$J/\psi$ channel, radiative decays can also help constrain the spin and nature of the observed structures. Radiative width ratios have already been used to probe the nature of $\chi_{c1}(3872)$~\cite{LHCb:2024tpv,Grinstein:2024rcu,Guo:2014taa}. The radiative-to-di-$J/\psi$ yield ratio provides a spin-sensitive constraint, while compatible pole positions extracted from both amplitudes would test a common resonance interpretation against a threshold enhancement \cite{Guo:2020pvt}.

Beyond the observed $2^{++}$ states, the $0^{++}$ and $1^{+-}$ sectors remain unestablished. Lattice quark-potential-model calculations~\cite{Zhang:2026LQPM} place the lightest $0^+$ and $1^+$ $cc\bar c\bar c$ states just above the $\eta_c\eta_c$ and $\eta_c J/\psi$ thresholds, respectively. Because these states lie so close to the thresholds, their strong decays are subject to threshold rescattering and to the broad $\eta_c$, which can obscure the resonance or turn it into a threshold enhancement. The radiative channels $\gamma J/\psi$ and $\gamma\eta_c$, with a photon in the final state, are not subject to the same hadronic final-state interactions and provide a complementary handle on the resonance. $C$-parity selects $0^{++}\to\gamma J/\psi$ and $1^{+-}\to\gamma\eta_c$, while the interchanged combinations are forbidden. The $\gamma J/\psi$ final state is shared with the $2^{++}$ states, but the $J/\psi$ dilepton angular distribution separates the scalar from the tensor. These two channels can therefore establish the $0^{++}$ and $1^{+-}$ states and determine their $J^{PC}$.

The $\gamma J/\psi$ and $\gamma\eta_c$ widths have so far been calculated only at LO \cite{Yang:2026bjy}. A complete one-loop calculation tests NRQCD factorization and calibrates the relation between radiative widths and tetraquark LDMEs. This is important at the charm scale, where NLO corrections to quarkonium annihilation and fully heavy tetraquark electromagnetic decays can reach tens of percent or order unity \cite{Barbieri:1979be,Petrelli:1997ge,Bodwin:2001pt,Liu:2025mxv}. The $0^{++}$ channel involves two $S$-wave color-spin operators and their interference, so its NLO hard coefficient is a $2\times2$ matrix, whereas the $2^{++}$ and $1^{+-}$ channels each have a single leading-$v$ operator. A common NLO normalization would preserve the LO color-spin hierarchy, whereas a nonuniform correction would change the relative weights and make the radiative width sensitive to the initial color-spin composition.

In this Letter, we present the first complete NLO QCD calculation of $T_{4c}(0^{++},2^{++})\to\gamma J/\psi$ and $T_{4c}(1^{+-})\to\gamma\eta_c$ at leading order in the relative velocity. We determine the short-distance coefficients for all relevant color-spin operators, including scalar interference, and predict the tensor $J/\psi$ polarization and dilepton angular distribution without an overall LDME normalization. For a spin-resolved tensor state, we show how the ratio of its $\gamma J/\psi$ and di-$J/\psi$ yields, combined with an inclusive-production fit and the total width, can constrain its LDME without a potential-model input. This extraction requires compatible production and decay normalizations. If scalar and tensor signals are unresolved, the same yield ratio can instead constrain their relative contributions, provided their respective radiative-to-hadronic decay ratios are known. We discuss these inputs and assumptions below.

\textit{NRQCD factorization and matching}. The decays $T_{4c}(0^{++},2^{++})\to\gamma J/\psi$ and $T_{4c}(1^{+-})\to\gamma\eta_c$ are treated in NRQCD at leading order in the heavy-quark relative velocity $v$ \cite{Bodwin:1994jh,Yang:2026bjy}. The $J/\psi$ and $\eta_c$ are projected onto the color-singlet ${}^3S_1^{[1]}$ and ${}^1S_0^{[1]}$ states, respectively. Their decay widths can be written as
\begin{equation}
\begin{aligned}
\Gamma_J={}&\frac{12\pi^4e_c^2\alpha\alpha_s^4(\mu_R)}{m_c^{11}}\rho_H\\
&\times\left[R_J^{(0)}+4\pi\alpha_s(\mu_R)R_J^{(1)}(\mu_R/m_c)\right],\\
R_J^{(i)}={}&\operatorname{Tr}\!\left[h_J^{(i)}\rho_{T,J}^{T}\right].
\end{aligned}
\label{eq:factorization}
\end{equation}
Here $\rho_H=\langle O_1^{J/\psi}({}^3S_1)\rangle$ for $J=0,2$ and $\rho_H=\langle O_1^{\eta_c}({}^1S_0)\rangle$ for $J=1$. The matrices $h_J^{(i)}$ and $\rho_{T,J}$ are the hard coefficients and tetraquark LDMEs. The NLO matrices $h_J^{(1)}$ follow from Born--one-loop interference in each color-spin channel.

For identical charm quarks, Pauli statistics restrict the allowed $S$-wave color-spin bases to
\begin{align}
\mathcal B_{0^{++}}={}&\left\{\big[(cc)_{\bar{\mathbf 3}}({}^3S_1)(\bar c\bar c)_{\mathbf 3}({}^3S_1)\big]_0,\right.\nonumber\\[-2pt]
&\left.\hspace{1.2em}\big[(cc)_{\mathbf 6}({}^1S_0)(\bar c\bar c)_{\bar{\mathbf 6}}({}^1S_0)\big]_0\right\},\nonumber\\
\mathcal B_{1^{+-},2^{++}}={}&\left\{\big[(cc)_{\bar{\mathbf 3}}({}^3S_1)(\bar c\bar c)_{\mathbf 3}({}^3S_1)\big]_{1,2}\right\}.
\label{eq:color-spin-bases}
\end{align}
For each operator $Q_n$, we define $L_n=\langle0|Q_n^{\rm ren}|T\rangle$ and $(\rho_T)_{nm}=L_nL_m^*$. The scalar hard coefficient and LDME are therefore $2\times2$ matrices. Their off-diagonal entries describe interference between the two color-spin amplitudes. The transpose in Eq.~\eqref{eq:factorization} follows from $|\sum_nF_nL_n|^2$ and is retained for complex LDMEs. The axial and tensor channels each contain one leading-$v$ operator. The tensor amplitude nevertheless has three physical polarization components, whose spin-averaged contraction gives a single hard coefficient. Reduced nonrelativistic state normalization is used throughout. State-conversion factors and initial-spin averages are included once in Eq.~\eqref{eq:factorization}.

The hard coefficients are determined by matching renormalized QCD amplitudes onto NRQCD with on-shell partonic states. The matching is performed at $P_T=P_H+k$, with $P_T^2=(4m_c)^2$, $P_H^2=(2m_c)^2$, and $k^2=0$. The potential-region contribution is isolated at finite relative velocity and subtracted before the zero-relative-momentum limit is taken \cite{Beneke:1997zp}. Dimensional regularization with $D=4-2\epsilon$ is used in both theories. The charm-quark mass and field are renormalized in the on-shell scheme, while $\alpha_s$ is renormalized in the $\overline{\rm MS}$ scheme with $n_f=4$. The nonpotential NRQCD loop contribution vanishes in pure dimensional regularization. Since $T_{4c}\to\gamma H g$ vanishes by color conservation, the NLO correction is purely virtual. The BMHV prescription is adopted for $\gamma_5$ in the axial channel \cite{tHooft:1972fi,Breitenlohner:1977hr}.

\begin{figure}[t]
\centering
\includegraphics[width=\columnwidth]{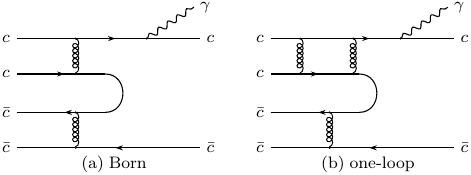}
\caption{Representative Feynman diagrams for $cc\bar c\bar c\to\gamma c\bar c$ at (a) Born and (b) one-loop order.}
\label{fig:radiative-feynman}
\end{figure}

We calculate the on-shell partonic amplitude $cc\bar c\bar c\to c\bar c+\gamma$ and project its external heavy-quark states onto the required color and spin configurations. Representative Born and one-loop diagrams are shown in Fig.~\ref{fig:radiative-feynman}. At Born level, our implementation also reproduces the short-distance coefficients for $T_{4c}\to\gamma\gamma$ using the same operator conventions~\cite{Sang:2023ncm}. The one-loop integrals reduce to 52 common master-integral classes. Their analytic finite combinations agree with independent numerical integral evaluations. The hard coefficients are finite after renormalization and matching. Photon Ward identities are verified before the transverse projection, and the $D$-dimensional spin factors are retained through the Laurent expansion. The logarithmic terms agree with the running of $\alpha_s^4$. The Supplemental Material provides operator conventions, finite amplitudes, matching checks, and numerical prescriptions \cite{SupplementalMaterial}.

\textit{NLO short-distance coefficients and radiative widths}. For $N_c=3$ and $n_f=4$, the hard coefficients are evaluated at the leading-$v$ matching point $M_T=4m_c$ and $M_H=2m_c$. The finite one-loop terms below correspond to $\mu_R=m_c$. In the ordered scalar basis $\{\bar{\mathbf 3}\otimes\mathbf 3,\mathbf 6\otimes\bar{\mathbf 6}\}$, the hard matrices are
\begin{equation}
\begin{aligned}
h_0^{(0)}={}&
\begin{pmatrix}
109.82991 & -13.03692\\
-13.03692 & 1.54750
\end{pmatrix},\\
h_0^{(1)}(1)={}&
\begin{pmatrix}
26.92184 & 4.32217-1.02746i\\
4.32217+1.02746i & -1.40542
\end{pmatrix}.
\end{aligned}
\label{eq:scalar-hard-results}
\end{equation}
The off-diagonal elements should be retained because they would contribute through $\rho_{36}$. More importantly, $h_0^{(1)}(1)$ is not proportional to $h_0^{(0)}$. The NLO correction therefore modifies the relative weights of the two color-spin directions and their interference differently. The positive correction to the $\bar{\mathbf 3}\otimes\mathbf 3$ diagonal coefficient and the negative correction to the $\mathbf 6\otimes\bar{\mathbf 6}$ coefficient make this color dependence explicit.

For the one-operator tensor and axial channels, the corresponding coefficients are
\begin{equation}
\begin{aligned}
\big(h_2^{(0)},h_2^{(1)}(1)\big)&=(10.30795,3.73544),\\
\big(h_1^{(0)},h_1^{(1)}(1)\big)&=(132.42274,48.34858),
\end{aligned}
\label{eq:single-hard-results}
\end{equation}
where no leading-$v$ color mixing occurs. Their tetraquark LDMEs therefore set the overall widths but cancel in the NLO-to-LO ratios.

The hard coefficients at a general renormalization scale are obtained from
\begin{equation}
h_J^{(1)}(r)=h_J^{(1)}(1)+\frac{\beta_0}{2\pi^2}\ln r\,h_J^{(0)},
\qquad r=\frac{\mu_R}{m_c},
\label{eq:hard-scale-running}
\end{equation}
where $\beta_0=(11N_c-2n_f)/3=25/3$. The central widths are evaluated at $\mu_R=2m_c$ using $h_J^{(1)}(2)$. For the numerical analysis, we take
\begin{equation}
\begin{aligned}
m_c=1.5~\mathrm{GeV}, e_c=\frac{2}{3}, \alpha=\frac{1}{137.065}, \alpha_s(2m_c)=0.249.
\end{aligned}
\label{eq:numerical-inputs}
\end{equation}
These inputs give $M_T=6.0$ GeV and $M_H=3.0$ GeV at the leading-$v$ matching point. Physical hadron masses differ by binding and relative-velocity effects. The production-based illustration below uses a separate mass and scale prescription.

The charmonium LDMEs are $\rho_{J/\psi}=1.32~\mathrm{GeV}^3$ and $\rho_{\eta_c}=0.44~\mathrm{GeV}^3$. The tetraquark LDMEs are taken from Refs.~\cite{Feng:2023agq,Yang:2026bjy,Celiberto:2025ziy} and listed in Table~\ref{tab:tetraquark-ldmes}. Table~\ref{tab:central-widths} gives the widths from Eq.~\eqref{eq:factorization}, their scale variations and charm-mass sensitivities, and $K_J=\Gamma_J^{\rm NLO}/\Gamma_J^{\rm LO}$.

\begin{table}[ht]
\caption{Tetraquark LDME inputs in $\mathrm{GeV}^9$. The scalar entries are ordered as $(\rho_{33},\rho_{36},\rho_{66})$.}
\label{tab:tetraquark-ldmes}
\scriptsize
\setlength{\tabcolsep}{3pt}
\begin{ruledtabular}
\begin{tabular}{lcc}
Channel & Model I & Model II\\
\hline
$0^{++}$ & $(0.0347,0.0211,0.0128)$ & $(0.0187,-0.0161,0.0139)$\\
$2^{++}$ & $0.0144$ & $0.0126$\\
$1^{+-}$ & $0.0260$ & $0.0160$\\
\end{tabular}
\end{ruledtabular}
\end{table}

\begin{table}[ht]
\caption{Radiative widths in MeV with separate scale and mass scans over $m_c\leq\mu_R\leq4m_c$ at the central mass and $m_c=1.5\pm0.1~\mathrm{GeV}$ at $\mu_R=2m_c$. Central widths and $K$ factors at $\mu_R=2m_c$ are rounded to two significant figures. Scan shifts are rounded sensitivities, not statistical errors.}
\label{tab:central-widths}
\setlength{\tabcolsep}{3pt}
\renewcommand{\arraystretch}{1.15}
\resizebox{\columnwidth}{!}{%
\begin{tabular}{llcccc}
\hline\hline
Channel & Input & LO (scale) & NLO (scale) & NLO (mass) & $K_J$\\
\hline
$0^{++}\to\gamma J/\psi$ & Model I & $0.73^{+1.72}_{-0.43}$ & $2.2^{+3.8}_{-1.2}$ & $2.2^{+3.0}_{-1.2}$ & $3.0$\\
& Model II & $0.55^{+1.31}_{-0.33}$ & $1.3^{+1.7}_{-0.7}$ & $1.3^{+1.8}_{-0.7}$ & $2.3$\\
$2^{++}\to\gamma J/\psi$ & Model I & $0.033^{+0.078}_{-0.020}$ & $0.10^{+0.18}_{-0.06}$ & $0.10^{+0.14}_{-0.06}$ & $3.0$\\
& Model II & $0.029^{+0.068}_{-0.017}$ & $0.088^{+0.16}_{-0.05}$ & $0.088^{+0.12}_{-0.05}$ & $3.0$\\
$1^{+-}\to\gamma\eta_c$ & Model I & $0.26^{+0.60}_{-0.15}$ & $0.78^{+1.4}_{-0.43}$ & $0.78^{+1.1}_{-0.44}$ & $3.1$\\
& Model II & $0.16^{+0.37}_{-0.09}$ & $0.48^{+0.86}_{-0.27}$ & $0.48^{+0.68}_{-0.27}$ & $3.1$\\
\hline\hline
\end{tabular}
}
\end{table}

At the central scale, the NLO widths in Table~\ref{tab:central-widths} are $2.3$--$3.1$ times the LO widths. The NLO corrections are therefore essential for quantitative estimates of these radiative transitions and for assessing their observability. We provide the corresponding NLO short-distance coefficients. The $\gamma J/\psi$ channel benefits from the clean $J/\psi\to\ell^+\ell^-$ signature, whereas the $\gamma\eta_c$ channel is limited by $\eta_c$ reconstruction: the useful hadronic branching fractions are at the few-percent level and the broad $32$~MeV $\eta_c$ width degrades the mass resolution. The gluon fragmentation function for the axial state vanishes at the input scale at LO. The forward calculation of Ref.~\cite{Celiberto:2025vra} predicts a lower production rate for the axial state than for the scalar and tensor states. A quantitative assessment requires production rates, total widths, reconstruction efficiencies, backgrounds, and mass resolution. 

\begin{figure}[t]
\centering
\includegraphics[width=\columnwidth]{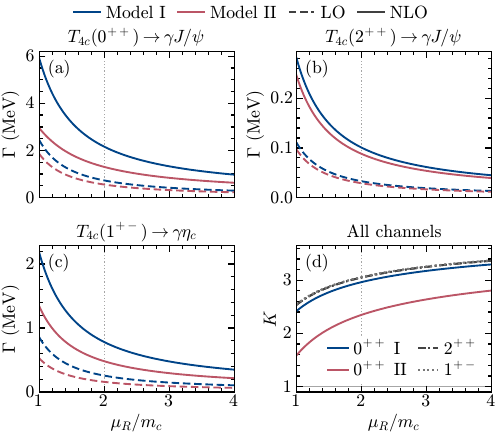}
\caption{Renormalization-scale dependence at $m_c=1.5$ GeV with fixed central LDMEs. (a--c) Scalar, tensor, and axial radiative widths. Navy and rose denote Models I and II, and dashed and solid lines denote LO and NLO. (d) $K=\Gamma_{\rm NLO}/\Gamma_{\rm LO}$. The tensor and axial $K$ factors are model independent and nearly coincide. Dotted vertical lines mark $\mu_R=2m_c$.}
\label{fig:scale-dependence}
\end{figure}

Table~\ref{tab:central-widths} also shows pronounced charm-mass and scale dependences, characteristic of slow perturbative convergence at the charm scale \cite{Barbieri:1979be,Petrelli:1997ge,Bodwin:2001pt,Feng:2017hlu}. At $m_c=1.4$ ($1.6$) GeV, all NLO widths are approximately $2.41$ ($0.44$) times their central values. The common $m_c^{-11}$ factor and running of $\alpha_s(2m_c)$ dominate the mass sensitivity, while $\alpha_s^4(\mu_R)$ and the LO-proportional logarithm govern scale variation. These predominantly common normalization changes do not discriminate color-spin configurations.

Figure~\ref{fig:scale-dependence} shows that the widths decrease monotonically with $\mu_R$, with milder slopes at larger scales as the running of $\alpha_s$ weakens. Over the scale scan, the ratio of the LO widths at the two endpoints is about $8.3$ and decreases to $6.1$--$6.3$ at NLO, or $4.7$ for scalar Model II. The NLO correction therefore improves the scale stability, although substantial residual dependence remains. Further stabilization may be possible with the optimized NRQCD strategy for exclusive double-$J/\psi$ production, which separates photon-fragmentation and nonfragmentation amplitudes and resums radiative and relativistic corrections with the measured $J/\psi$ decay constant \cite{Sang:2023liy}.

The $K$ factors in Table~\ref{tab:central-widths} and Figure~\ref{fig:scale-dependence} reveal a channel-dependent pattern. In the $2^{++}$ and $1^{+-}$ channels, the Model I and Model II $K$ factors coincide because the single leading-$v$ LDME cancels in $\Gamma_{\rm NLO}/\Gamma_{\rm LO}$, whereas the scalar $K$ factors remain distinct because the two color components and their interference receive different NLO corrections.

\textit{Tensor polarization}. The tensor amplitude provides an observable independent of the overall LDME normalization: the longitudinal fraction $f_L=\Gamma_L/\Gamma$ of the daughter $J/\psi$. In a physical three-component basis, the longitudinal and transverse polarization metrics are $G_L=\operatorname{diag}(0,0,2/5)$ and $G_T=\operatorname{diag}(4/5,2/5,0)$. The common tetraquark and charmonium LDMEs, phase space, and $m_c^{-11}$ factor cancel in the fraction. Writing $\Gamma=\Gamma^{(0)}+\Delta\Gamma$ and similarly for $\Gamma_L$, the strictly expanded NLO result is
\begin{equation}
f_L^{\rm NLO}=f_L^{(0)}
+\frac{\Delta\Gamma_L-f_L^{(0)}\Delta\Gamma}{\Gamma^{(0)}},
\qquad
f_L^{(0)}=\frac{\Gamma_L^{(0)}}{\Gamma^{(0)}}.
\label{eq:polarization-expanded}
\end{equation}
At $\mu_R=2m_c$, we obtain $f_L^{(0)}=0.11$ and $f_L^{\rm NLO}=0.18$. The unexpanded quotient $(\Gamma_L^{(0)}+\Delta\Gamma_L)/(\Gamma^{(0)}+\Delta\Gamma)$ instead gives $0.14$. The two prescriptions agree through NLO, but differ by $0.047$ at the central scale because $\Delta\Gamma/\Gamma^{(0)}\simeq2.05$. This difference diagnoses higher-order sensitivity. It does not bound the unknown higher-order terms.

\begin{figure}[t]
\centering
\includegraphics[width=\columnwidth]{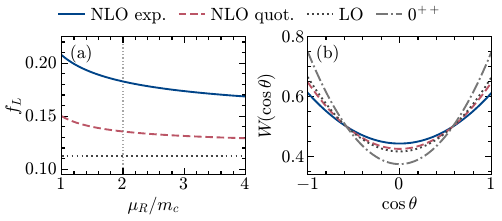}
\caption{Polarization in $T_{4c}(2^{++})\to\gamma J/\psi$. (a) Longitudinal fraction versus renormalization scale. (b) Normalized dilepton distribution at $\mu_R=2m_c$. The curves show tensor LO, strictly expanded NLO (exp.), and the quotient of NLO widths (quot.). The scalar reference appears only in (b). The two NLO prescriptions illustrate higher-order sensitivity.}
\label{fig:polarization}
\end{figure}

In Fig.~\ref{fig:polarization}(a), the LO fraction is scale independent. The expanded NLO curve remains above the quotient throughout $m_c\leq\mu_R\leq4m_c$. Their central separation exceeds either curve's change to a scale endpoint, so scale variation alone misses this higher-order sensitivity. The universal renormalization logarithm cancels in $f_L$, whose remaining NLO scale dependence enters through $\alpha_s$.

The polarization fraction can be accessed through $J/\psi\to\ell^+\ell^-$. Let $\theta$ be the angle of $\ell^+$ in the $J/\psi$ rest frame relative to the $J/\psi$ direction in the tetraquark rest frame. After integration over the lepton azimuth, the normalized distribution is \cite{Faccioli:2010kd}
\begin{equation}
W(z)\equiv\frac{1}{\Gamma_{\ell\ell}}
\frac{\mathrm d\Gamma_{\ell\ell}}{\mathrm dz}
=\frac38\big[(1+f_L)+(1-3f_L)z^2\big],
\quad z=\cos\theta.
\label{eq:lepton-angle}
\end{equation}
Here $\Gamma_{\ell\ell}$ denotes the cascade rate for $T_{4c}\to\gamma J/\psi\to\gamma\ell^+\ell^-$. For a scalar parent and a real photon, angular momentum conservation requires $f_L=0$, independently of the scalar color mixture.

Figure~\ref{fig:polarization}(b) compares the resulting angular shapes at the central scale. The scalar reference, $W(z)=3(1+z^2)/8$, has the strongest rise toward $|z|=1$. A nonzero tensor $f_L$ increases the distribution near $z=0$ and reduces it near $|z|=1$, making the tensor shape flatter. The strictly expanded NLO result gives the largest flattening. The quotient prediction lies between it and the tensor LO curve. Thus the angular distribution supplies spin information without strong-decay inputs, while the separation between the two NLO tensor curves shows the present perturbative limitation of a quantitative comparison. These shapes assume an unpolarized parent, or integration over its complete decay solid angle. A fiducial measurement must correct for polarization-dependent acceptance or fit the parent spin density matrix jointly with the decay angles.

\textit{Tensor LDME extraction}. Assuming a pure-$2^{++}$ state $X$, the radiative width retains its LDME dependence. The radiative yield measures $\sigma_X\mathcal B_{\gamma\psi}=\sigma_X\Gamma_{\gamma\psi}/\Gamma_{\rm tot}$. Comparing radiative and di-$J/\psi$ decays cancels the production normalization,
\begin{equation}
R_{\gamma/\psi\psi}
\equiv
\frac{\mathcal B(X\to\gamma J/\psi)}
{\mathcal B(X\to J/\psi J/\psi)}
=
\frac{\Gamma(X\to\gamma J/\psi)}
{\Gamma(X\to J/\psi J/\psi)}.
\label{eq:radiative-ratio}
\end{equation}
The production cross section, luminosity, and total width cancel in this ratio, but the unknown di-$J/\psi$ branching fraction prevents extracting $\rho_{T,2}^{\rm fix}$ from this ratio alone.

The complementary input is provided by the NLO+NLL analysis of inclusive production, $pp\to T_{4c}(2^{++})+X_{\rm incl}$ \cite{Wang:2025hex}. The di-$J/\psi$ data constrain
\begin{equation}
\begin{aligned}
P_2^{\rm fit}&=\rho_{T,2}^{\rm prod}\mathcal B_{\psi\psi}
=5\xi\rho_{T,2}^{\rm fix}\mathcal B_{\psi\psi},\\
P_2^{\rm fit}&=\left(2.22\pm0.80{}^{+1.29}_{-0.57}
{}^{+0.62}_{-0.00}\right)\times10^{-4}~\mathrm{GeV}^9.
\end{aligned}
\label{eq:production-constraint}
\end{equation}
The three uncertainties arise from experiment, scale variation, and PDFs. Here $\rho_{T,2}^{\rm prod}$ denotes the fitted production-convention LDME. At leading order in $v$ and under vacuum saturation, $5\xi$ relates it to the fixed-polarization decay LDME---the factor $5$ sums the $2J+1=5$ tensor polarizations---including operator, state, and polarization conventions. Identical currents and states with a spin-summed production LDME give $\xi=1$. This identification remains unverified for the external fit. The initial-spin average is already in the decay coefficient.

Writing $\Gamma_{\gamma\psi}=H_2^{\rm NLO}\rho_{T,2}^{\rm fix}$ and $\Gamma_{\psi\psi}=\mathcal B_{\psi\psi}\Gamma_{\rm tot}$, the two constraints give
\begin{equation}
\rho_{T,2}^{\rm fix}
=\sqrt{\frac{P_2^{\rm fit}R_{\gamma/\psi\psi}\Gamma_{\rm tot}}
{5\xi H_2^{\rm NLO}}}.
\label{eq:ldme-extraction}
\end{equation}
This relation avoids a potential-model tetraquark LDME input. For illustration, match the production mass and central hard scale, $M=6.847$ GeV, $m_c=M/4$, and $\mu_R=M$ ($\mu_F=M$ for production) \cite{Wang:2025hex}, together with $P_2^{\rm fit}=2.22\times10^{-4}~\mathrm{GeV}^9$ and the CMS interference-fit width $\Gamma_{\rm tot}=135~\mathrm{MeV}$ \cite{CMS:2026tiu}. Then
\begin{equation}
\rho_{T,2}^{\rm fix}\simeq
0.099\sqrt{R_{\gamma/\psi\psi}/\xi}~\mathrm{GeV}^9.
\end{equation}
This coefficient uses our stated decay inputs. Their compatibility with the production fit and $\xi$ remain conditions. Without fixing $\xi$, only the scaled combination $\sqrt\xi\,\rho_{T,2}^{\rm fix}$ is determined; the quoted CMS width errors propagate to about $8\%$ on $\rho_{T,2}^{\rm fix}$ through the square root~\cite{CMS:2026tiu}. A mixed sample requires spin-resolved yields and a compatible production fit~\cite{Celiberto:2026kks}.

\textit{Disentanglement of scalar and tensor contributions}. For unresolved $0^{++}$ and $2^{++}$ signals in one mass window, the di-$J/\psi$ spectrum gives their combined yield. The radiative channel constrains their fractions through distinct radiative-to-hadronic responses.

For the inputs of Table~\ref{tab:tetraquark-ldmes} at $m_c=1.5$ GeV and $\mu_R=2m_c$, $\gamma\gamma$ is suppressed by order $10^2$ or more relative to $\gamma J/\psi\to\gamma\ell^+\ell^-$, taking $\mathcal B(J/\psi\to\ell^+\ell^-)\simeq0.06$ for either charged lepton \cite{Sang:2023ncm,Liu:2025mxv}. Tensor $\eta_c\eta_c$ requires a $D$ wave and lacks a reliable width \cite{Anwar:2017toa,Chen:2022sbf}, limiting these alternatives for quantitative spin separation.

Assuming that the resonant yield in the selected mass window is saturated by $0^{++}$ and $2^{++}$ contributions, define the scalar fraction of the di-$J/\psi$ signal by
\begin{equation}
f_0^{\psi\psi}
\equiv
\frac{\sigma_0\mathcal B_0(\psi\psi)}{\sigma_0\mathcal B_0(\psi\psi)+\sigma_2\mathcal B_2(\psi\psi)},
\qquad
f_2^{\psi\psi}=1-f_0^{\psi\psi},
\label{eq:spin-fraction}
\end{equation}
where $\sigma_J$ denotes the production cross section in the same fiducial region. For the pure-spin responses $r_J^{\gamma\psi}\equiv\Gamma_J(\gamma J/\psi)/\Gamma_J(J/\psi J/\psi)$, one has $\mathcal B_J(\gamma J/\psi)=r_J^{\gamma\psi}\mathcal B_J(J/\psi J/\psi)$. For this unresolved sample, $R_{\gamma/\psi\psi}$ denotes the combined yield ratio,
\begin{equation}
\begin{aligned}
R_{\gamma/\psi\psi}
&=\frac{\sigma_0\mathcal B_0(\gamma\psi)+\sigma_2\mathcal B_2(\gamma\psi)}
{\sigma_0\mathcal B_0(\psi\psi)+\sigma_2\mathcal B_2(\psi\psi)}\\
&=r_0^{\gamma\psi}f_0^{\psi\psi}+r_2^{\gamma\psi}f_2^{\psi\psi}\\
&=r_2^{\gamma\psi}+(r_0^{\gamma\psi}-r_2^{\gamma\psi})f_0^{\psi\psi}.
\end{aligned}
\label{eq:spin-response}
\end{equation}
If $r_0^{\gamma\psi}\neq r_2^{\gamma\psi}$, this relation can be inverted directly:
\begin{equation}
f_0^{\psi\psi}
=\frac{R_{\gamma/\psi\psi}-r_2^{\gamma\psi}}{r_0^{\gamma\psi}-r_2^{\gamma\psi}},
\qquad
f_2^{\psi\psi}=1-f_0^{\psi\psi}.
\label{eq:spin-inversion}
\end{equation}
The response coefficients require external input for the strong di-$J/\psi$ widths. The inversion is therefore conditional on the strong-decay model and the state assignment, with the radiative and strong widths in each response describing the same state. Alternatively, let $g_2^{\gamma\psi}$ denote the tensor fraction of the efficiency-corrected radiative yield in the same mass window and production region. Since the scalar has $f_{L,0}=0$ and Eq.~\eqref{eq:lepton-angle} is linear in $f_L$, an incoherent two-spin sample satisfies
\begin{equation}
f_{L,\rm mix}=g_2^{\gamma\psi}f_{L,2},
\qquad
g_2^{\gamma\psi}=\frac{f_{L,\rm mix}}{f_{L,2}}.
\label{eq:polarization-mixture}
\end{equation}
This determines the radiative tensor fraction without strong-decay inputs, subject to the tensor-polarization prediction and acceptance conditions above. It does not by itself determine the di-$J/\psi$ spin fractions.

Eq.~\eqref{eq:spin-response} treats the scalar and tensor components as incoherent yields. If the selected mass window contains overlapping spin amplitudes, additional partial waves, or an interfering continuum background, the corresponding interference terms must be included in a joint angular and lineshape analysis.

\textit{Conclusion}. We have calculated the complete NLO QCD corrections to three fully charm tetraquark radiative transitions at leading $v$. The NLO scalar hard matrix is not proportional to its LO counterpart, changing the relative contributions of the color-spin components and their interference. For the chosen LDMEs at $m_c=1.5$ GeV and $\mu_R=2m_c$, the NLO widths are $2.3$--$3.1$ times the LO widths.

The dilepton angular distribution probes the parent spin without an overall LDME normalization, and NLO corrections affect this spin-sensitive observable. At $\mu_R=2m_c$, the tensor longitudinal fraction $f_L$ changes from $0.11$ at LO to $0.18$ at strictly expanded NLO, although the prediction remains sensitive to higher-order corrections. The NLO-corrected radiative widths provide updated inputs for constraining tensor LDMEs and scalar--tensor fractions in combination with compatible production and strong-decay information. Our results provide important theoretical input for constraining the properties of fully charm tetraquarks and for interpreting the near-threshold $0^{++}$, $1^{+-}$, and $2^{++}$ resonances.

\textit{Acknowledgments}. The authors thank Long-Bin Chen, Qi-Ming Feng, and Hao Yang for useful discussions. This work is supported by the National Natural Science Foundation of China (NSFC) under Grants No. 12547114, 12325503, and 12547172, and by the High-Performance Computing Platform of Peking University. X. L. is also supported by the China Postdoctoral Science Foundation under Grant No. 8206301723.
\newpage
\bibliography{ref}

\onecolumngrid
\newpage
\input{supplement.tex}

\end{document}

%% file: supplement.tex
\setcounter{equation}{0}\setcounter{table}{0}\setcounter{figure}{0}\setcounter{section}{0}
\renewcommand{\theequation}{S\arabic{equation}}
\renewcommand{\thetable}{S\Roman{table}}
\renewcommand{\thefigure}{S\arabic{figure}}
\renewcommand{\thesection}{S\arabic{section}}
\renewcommand{\theHsection}{supp.\arabic{section}}
\renewcommand{\theHequation}{supp.\arabic{equation}}
\renewcommand{\theHtable}{supp.\arabic{table}}
\renewcommand{\theHfigure}{supp.\arabic{figure}}
\setcounter{secnumdepth}{2}
\setlength{\abovedisplayskip}{6pt plus 2pt minus 2pt}
\setlength{\belowdisplayskip}{6pt plus 2pt minus 2pt}
\setlength{\jot}{2pt}
\setlength{\textfloatsep}{10pt plus 2pt minus 2pt}
\setlength{\intextsep}{10pt plus 2pt minus 2pt}
\raggedbottom
\begin{center}
{\large\bfseries Supplemental Material}\\[4pt]
{\bfseries Next-to-leading order QCD corrections to fully charm tetraquark radiative decays}\\[4pt]
Xiang Li, Yan-Qing Ma, Bo-Xuan Shi, and Sheng-Qi Zhang
\end{center}

\section{Operator conventions and matching}
\label{sec:factorization}
With $D=4-2\epsilon$, $d_s=D-1$, and two-component heavy-quark fields $\psi,\chi$, define
\begin{align}
C_{\bar3}^{ab,cd}&=\frac{\delta^{ac}\delta^{bd}-\delta^{ad}\delta^{bc}}{2\sqrt3}, &
C_6^{ab,cd}&=\frac{\delta^{ac}\delta^{bd}+\delta^{ad}\delta^{bc}}{2\sqrt6},\nonumber\\
S_0^{ab}&=\psi_a^T i\sigma^2\psi_b, & S_1^{ab,i}&=\psi_a^T i\sigma^2\sigma^i\psi_b,\nonumber\\
\bar S_0^{cd}&=\chi_c^\dagger i\sigma^2\chi_d^*, & \bar S_1^{cd,j}&=\chi_c^\dagger\sigma^j i\sigma^2\chi_d^*.
\end{align}
The physical scalar phases and axial current are fixed by
\begin{equation}
Q_{\bar3}=-\frac{\delta_{ij}}{\sqrt{d_s}}S_1^{ab,i}\bar S_1^{cd,j}C_{\bar3}^{ab,cd},\qquad
Q_6=S_0^{ab}\bar S_0^{cd}C_6^{ab,cd},\qquad
Q_1^i=-\frac{i\epsilon^{ijk}}{\sqrt2}S_1^{ab,j}\bar S_1^{cd,k}C_{\bar3}^{ab,cd}.
\end{equation}
For each normalized symmetric traceless tensor polarization $\epsilon^{ij}(\lambda)$, define
\begin{equation}
Q_2(\lambda)=\epsilon_{ij}^*(\lambda)S_1^{ab,i}\bar S_1^{cd,j}C_{\bar3}^{ab,cd},
\qquad \epsilon_{ij}^*(\lambda)\epsilon^{ij}(\lambda')=\delta_{\lambda\lambda'}.
\label{eq:tensor-current}
\end{equation}
The spin-two projector is $\Pi^{ij,kl}=\tfrac12(\delta^{ik}\delta^{jl}+\delta^{il}\delta^{jk})-\delta^{ij}\delta^{kl}/d_s$. Axial external polarizations remain four-dimensional in BMHV. Loop momenta and contracted QCD indices are $D$-dimensional~\cite{tHooft:1972fi,Breitenlohner:1977hr}. We use $L_n=\langle0|Q_n^{\rm ren}|T_{\rm rel}\rangle/\sqrt{2M_T}$ and $\rho_{nm}=L_nL_m^*$, with spinful tetraquark decay LDMEs defined for one polarization. The charmonium convention is $\rho_\psi=\rho_{J/\psi}$ (spin summed) and $\rho_\eta=\rho_{\eta_c}$.

Matching uses the same labelled on-shell partonic amplitude and operator matrix elements $\Phi$, fixing identical-particle factors. With coupling factors included,
\begin{equation}
\mathcal M^{(0)}=C^{(0)}\Phi^{(0)},\qquad
C^{(1)}\Phi^{(0)}=\mathcal M_{\rm QCD,ren}^{(1)}-C^{(0)}\Phi_{\rm EFT,ren}^{(1)}.
\label{eq:partonic-matching}
\end{equation}
The common Coulomb term is subtracted at finite relative velocity with kinetic propagators retained, before the zero-velocity hard limit~\cite{Beneke:1997zp}. Nonpotential integrals are scaleless, with ultraviolet and infrared poles tracked separately. The half-LSZ residues arise by differentiating the heavy-quark self-energy with respect to a residual energy before the on-shell limit. In Feynman gauge, $a_s=\alpha_s/(4\pi)$ gives
\begin{equation}
\left.\Phi_{\rm np}^{(1)}\right|_{\rm pairs}
=-8a_s\left(\frac1{\epsilon_{\rm UV}}-\frac1{\epsilon_{\rm IR}}\right)\Phi^{(0)}
=-\left.\Phi_{\rm np}^{(1)}\right|_{\rm half\text{-}LSZ}.
\end{equation}
Thus $Z_{PP}^{(1)}=0$ and the on-shell nonpotential matrix element vanishes. Leading density vertices preserve the Pauli tensor. Subleading terms supply no derivative-free $v^0$ conversion. Dimensional spin factors are retained through pole--finite products~\cite{Petrelli:1997ge}. Different residual-energy or evanescent conventions require finite coefficient conversion.

The workflow uses FeynArts~\cite{Hahn:2000kx}, CalcLoop~\cite{Ma:2023calcloop}, Kira~\cite{Klappert:2020nbg}, and AMFlow~\cite{Liu:2022chg}. With the renormalization schemes stated in the Letter, $\mathcal M_{\rm ct}=m_c\delta Z_m\mathcal M_{m,\rm unit}+(3\delta Z_2+4\delta Z_g)\mathcal M^{(0)}$, where $\mathcal M_{m,\rm unit}$ sums unit internal mass insertions. Our diphoton LO short-distance coefficients agree with the corrected arXiv v3 results of Ref.~\cite{Sang:2023ncm}. The LO radiative amplitudes of Ref.~\cite{Yang:2026bjy} are twice ours, and its widths are four times ours at fixed LDMEs, consistent with the pre-revision normalization of Ref.~\cite{Sang:2023ncm}; its scalar numbers additionally apply mixing-angle weights to the density entries.

\par
\clearpage
\begingroup
\allowdisplaybreaks[4]
\setlength{\jot}{2pt}
\setlength{\abovedisplayskip}{5pt plus 1pt minus 1pt}
\setlength{\belowdisplayskip}{5pt plus 1pt minus 1pt}
\section{Amplitudes, elementary functions, and dilogarithm combinations}
\label{sec:definitions}
We collect here the analytic ingredients that enter the finite one-loop coefficients.
The hard coefficients and amplitudes below are evaluated at the leading-$v$ matching point
$M_T=4m_c$, $M_H=2m_c$. The two scalar color components, the axial amplitude, and the three
tensor polarization components are ordered as
\[
a=\bigl((0,\bar3),(0,6),1,(2,1),(2,2),(2,3)\bigr),
\]
and this ordering is used throughout for all component-indexed quantities below.

Let $\ell_n=\ln n$, $L=\ln(2+\sqrt3)$, $A=(\pi-\arctan\sqrt{15})/\sqrt{15}$, and $U=(\ell_{19}-\ell_5)(\ell_{19}+\ell_{11})+4\ell_{19}(\ell_2-\ell_3)$. The ordered bases are

\begin{equation}
\begin{aligned}
(X_{1},\ldots,X_{9})={}&\bigl(1,\pi^2,\ell_2,\ell_2\ell_{11},\ell_2^2,\ell_2\ell_{23},\ell_3,\ell_2\ell_3,\ell_3^2\bigr),\\
(X_{10},\ldots,X_{18})={}&\bigl(\ell_5,\ell_2\ell_5,\ell_3\ell_5,\ell_5^2,\ell_7\ell_{11},\ell_2\ell_7,\ell_3\ell_7,\ell_5\ell_7,\ell_7^2\bigr),\\
(X_{19},\ldots,X_{26})={}&\bigl(L/\sqrt3,L^2,A,\ell_{11}(\ell_{11}-\ell_3),U,\ell_{23}(\ell_{23}-\ell_5),L(\ell_2+5\ell_3),\ell_3\ell_{23}\bigr),\\
(Y_1,\ldots,Y_8)={}&(1,1/\sqrt3,\ell_{19},\ell_2,\ell_3,\ell_5,\ell_7,L).
\end{aligned}
\label{eq:elementary-bases}
\end{equation}

All exact elementary terms are then specified compactly by
\begin{equation}
P_a=\frac1{D_a}\sum_{j=1}^{26}p_{ja}X_j,\qquad
R_a=\frac1{E_a}\sum_{j=1}^{8}r_{ja}Y_j,
\label{eq:finite-basis}
\end{equation}
with integer coefficients in Table~\ref{tab:elementary-weights}.

\begin{table}[!ht]
\caption{Exact integer coefficients for the elementary terms in Eq.~\eqref{eq:finite-basis}. The upper and lower blocks specify $P_a$ and $R_a$, respectively.}
\label{tab:elementary-weights}
\begingroup
\small
\setlength{\tabcolsep}{2pt}
\begin{ruledtabular}
\renewcommand{\arraystretch}{0.95}
\begin{tabular}{crrrrrr}
 & \multicolumn{2}{c}{$0^{++}$} & $1^{+-}$ & \multicolumn{3}{c}{$2^{++}$} \\
$a$ & $(0,\bar3)$ & $(0,6)$ & $1$ & $(2,1)$ & $(2,2)$ & $(2,3)$ \\
\hline
$D_a$ & $36050130436320$ & $698788490400$ & $62063451450$ & $21858371458560$ & $233709643440$ & $2214724861440$ \\
\hline
$p_{1a}$ & $153382481860608$ & $3376212840000$ & $249895435293504$ & $65803026167424$ & $286648873056$ & $4776634437120$ \\
$p_{2a}$ & $-21950333935915$ & $-2071816454785$ & $2701134605785$ & $12620920277315$ & $409324602785$ & $3093351685417$ \\
$p_{3a}$ & $31160745441540$ & $15699772145772$ & $64738227616920$ & $29841726030480$ & $-2675758412340$ & $-6056591851440$ \\
$p_{4a}$ & $-97069140630$ & $159734785518$ & $-521687454750$ & $-982460668770$ & $-7577557470$ & $-53012318058$ \\
$p_{5a}$ & $358540335047265$ & $8576334510891$ & $411830460506745$ & $140959588915635$ & $-3918710259345$ & $-138175333233$ \\
$p_{6a}$ & $50606556000$ & $-253032780000$ & $1177147171200$ & $1756730976000$ & $37426389000$ & $248052604800$ \\
$p_{7a}$ & $29484105712776$ & $7081359250680$ & $22011874235208$ & $-22033212993792$ & $-1564039533888$ & $-5348790967104$ \\
$p_{8a}$ & $-353200080769275$ & $1414234050783$ & $-317451527815395$ & $-17423041960605$ & $4980330053430$ & $15229938854451$ \\
$p_{9a}$ & $109130009964645$ & $2085058266087$ & $73718304296085$ & $-22625845125225$ & $-2423466979035$ & $-10404901160877$ \\
$p_{10a}$ & $-7357682372040$ & $453511978920$ & $-24848563259520$ & $-13016545607520$ & $393449385420$ & $2223416744640$ \\
$p_{11a}$ & $-23028142805370$ & $-4803782759550$ & $-68285977265130$ & $-29556208580670$ & $129422219130$ & $2095462501578$ \\
$p_{12a}$ & $-17590076904030$ & $-4030100487690$ & $-41349932780670$ & $-14842201781130$ & $-297188692530$ & $20094924462$ \\
$p_{13a}$ & $3014354515320$ & $352185744600$ & $12084634947780$ & $1123490240820$ & $-273631059630$ & $-1330740869868$ \\
$p_{14a}$ & $-35332889130$ & $-148946471982$ & $521687454750$ & $269580005730$ & $2029848030$ & $128461166442$ \\
$p_{15a}$ & $17703941138130$ & $-1731933204378$ & $6074119979850$ & $8108498305830$ & $-151644853230$ & $-2097586546770$ \\
$p_{16a}$ & $5706481546650$ & $-320336980098$ & $579208291650$ & $2639625287070$ & $-28011973830$ & $-952688179434$ \\
$p_{17a}$ & $-10121311200$ & $50606556000$ & $-215658643200$ & $-351346195200$ & $-7485277800$ & $-49610520960$ \\
$p_{18a}$ & $-4021935598110$ & $505737007638$ & $4516781337990$ & $1191048769230$ & $22143257340$ & $-193276551750$ \\
$p_{19a}$ & $118540681459200$ & $-1258636579200$ & $276808101570000$ & $127544502430800$ & $6851080618200$ & $24832847697840$ \\
$p_{20a}$ & $-80307254152125$ & $-4332548235015$ & $-161583030173565$ & $-76435784190675$ & $-654510524850$ & $-6873191738595$ \\
$p_{21a}$ & $40745397200508$ & $-18707035238892$ & $42824884526424$ & $-89872368820416$ & $-113211808164$ & $-10721585111328$ \\
$p_{22a}$ & $66201014880$ & $-5394156768$ & $0$ & $356440331520$ & $2773854720$ & $-37724424192$ \\
$p_{23a}$ & $-30868125750$ & $154340628750$ & $-521687454750$ & $-626020337250$ & $-4803702750$ & $-90736742250$ \\
$p_{24a}$ & $-10121311200$ & $50606556000$ & $-248609961600$ & $-351346195200$ & $-7485277800$ & $-49610520960$ \\
$p_{25a}$ & $15084185591475$ & $636568958025$ & $26510512543515$ & $9365702980725$ & $87820673850$ & $1009341803925$ \\
$p_{26a}$ & $0$ & $0$ & $65902636800$ & $0$ & $0$ & $0$ \\
\hline
$E_a$ & $273107048760$ & $5293852200$ & $940355325$ & $1821530954880$ & $19475803620$ & $184560405120$ \\
\hline
$r_{1a}$ & $40184646216$ & $-3942615144$ & $-16898747424$ & $238155376368$ & $2361023067$ & $-111045350208$ \\
$r_{2a}$ & $1191890700$ & $108353700$ & $-12383280000$ & $-374470387200$ & $-49409287200$ & $6241173120$ \\
$r_{3a}$ & $467698875$ & $-2338494375$ & $15808710750$ & $104336722875$ & $800617125$ & $15122790375$ \\
$r_{4a}$ & $115310738360$ & $-7972032856$ & $578403200880$ & $2193951493120$ & $-32924924180$ & $-211817477464$ \\
$r_{5a}$ & $4385745$ & $-21928725$ & $29239798770$ & $148109446485$ & $1855944090$ & $29243889675$ \\
$r_{6a}$ & $-30376508000$ & $6173733640$ & $-19479349800$ & $-595837838500$ & $4696376000$ & $-89833764050$ \\
$r_{7a}$ & $-27445832960$ & $869989120$ & $-195479298560$ & $-775102237000$ & $2238347930$ & $151208083936$ \\
$r_{8a}$ & $-1816214400$ & $-165110400$ & $8255520000$ & $208039104000$ & $22537569600$ & $3120586560$ \\
\end{tabular}
\end{ruledtabular}
\endgroup
\end{table}

The dilogarithm entries are
\begin{equation}
\begin{gathered}
\Phi_j=\operatorname{Re}\operatorname{Li}_2(z_j)\qquad\text{(principal branch, Feynman }+i0\text{)},\\[2pt]
\begin{aligned}
(z_{1},\ldots,z_{9})&=\left(\frac18,\frac3{23},\frac3{22},\frac17,\frac4{27},\frac4{25},\frac16,\frac5{27},\frac9{44}\right),\qquad &(z_{10},\ldots,z_{18})&=\left(\frac5{24},\frac3{14},\frac5{23},\frac14,\frac3{11},\frac9{32},\frac3{10},\frac{19}{63},\frac13\right),\\
(z_{19},\ldots,z_{25})&=\left(\frac{11}{32},\frac{19}{54},\frac{23}{63},\frac9{23},\frac25,\frac37,\frac9{19}\right),\qquad &(z_{26},\ldots,z_{29})&=\left(\frac73-\frac4{\sqrt3},\frac23-\frac1{\sqrt3},21-12\sqrt3,3\sqrt3-5\right),
\end{aligned}\\[2pt]
(z_{30},\ldots,z_{36})=\left(\frac{50}{95+9i\sqrt{15}},\frac{140}{95+9i\sqrt{15}},\frac{20}{5+3i\sqrt{15}},\frac{10}{5+i\sqrt{15}},\frac{20}{25+i\sqrt{15}},\frac{30}{25+i\sqrt{15}},-\frac{10}{5+3i\sqrt{15}}\right).
\end{gathered}
\end{equation}

The combinations entering the coefficients are
\begin{equation}
\begin{aligned}
(\Psi_1,\ldots,\Psi_9)&=(\Phi_1,\Phi_4,\Phi_6,\Phi_7,\Phi_{13},\Phi_{16},\Phi_{18},\Phi_{23},\Phi_{24}),\\
\Psi_{10}&=\Phi_3+\Phi_8-\Phi_9+\Phi_{10}-\Phi_{17}-\Phi_{20}-2\Phi_{25},\\
\Psi_{11}&=\Phi_{12}+\Phi_{15},\quad \Psi_{12}=-\Phi_{21}+\Phi_{22},\quad \Psi_{13}=\Phi_{11}+2\Phi_{14}-\Phi_{19},\\
\Psi_{14}&=\Phi_{26}-\Phi_{27}-\Phi_{28}-\Phi_{29},\quad \Psi_{15}=-\Phi_{30}+\Phi_{31},\quad \Psi_{16}=\Phi_{32}-\Phi_{36},\\
\Psi_{17}&=\Phi_{33},\quad \Psi_{18}=\Phi_{34}-\Phi_{35},\quad \Psi_{19}=\Phi_2+\Phi_5.
\end{aligned}
\label{eq:dilog-combinations}
\end{equation}

The finite coefficients are
\begin{equation}
c_a=s_a\left[P_a+i\pi R_a+\frac1{N_a}\sum_{k=1}^{19}b_{ka}\Psi_k\right],
\qquad
s_a=(1,1,135/140608,1,1,1).
\label{eq:finite-amplitude}
\end{equation}
with $N_a$ and the integer weights $b_{ka}$ in Table~\ref{tab:dilog-weights}.

\begin{table}[!ht]
\caption{Exact integer weights of the dilogarithm combinations in Eq.~\eqref{eq:finite-amplitude}, in the same component order as Table~\ref{tab:elementary-weights}.}
\label{tab:dilog-weights}
\begingroup
\small
\setlength{\tabcolsep}{2pt}
\begin{ruledtabular}
\renewcommand{\arraystretch}{0.95}
\begin{tabular}{crrrrrr}
 & \multicolumn{2}{c}{$0^{++}$} & $1^{+-}$ & \multicolumn{3}{c}{$2^{++}$} \\
$a$ & $(0,\bar3)$ & $(0,6)$ & $1$ & $(2,1)$ & $(2,2)$ & $(2,3)$ \\
\hline
$N_a$ & $1201671014544$ & $116464748400$ & $2068781715$ & $728612381952$ & $7790321448$ & $369120810240$ \\
\hline
$b_{1a}$ & $-190479669336$ & $48985790040$ & $-23714399808$ & $35074954368$ & $2817185508$ & $145466456256$ \\
$b_{2a}$ & $-75178504973$ & $46827422213$ & $384001290959$ & $312367160209$ & $1669261087$ & $-165785097959$ \\
$b_{3a}$ & $675681188$ & $-16892029700$ & $21586835634$ & $63332394944$ & $652774031$ & $21856521050$ \\
$b_{4a}$ & $583588315482$ & $740372947590$ & $1346671776882$ & $523668196014$ & $9982939758$ & $-29809743582$ \\
$b_{5a}$ & $5535310622767$ & $-179021864863$ & $5278330821273$ & $1144961541037$ & $-38422925516$ & $-225633278099$ \\
$b_{6a}$ & $-1013985336$ & $25349633400$ & $-35985049828$ & $-114953250048$ & $-1056038802$ & $-35444621940$ \\
$b_{7a}$ & $6693482986576$ & $-88786135936$ & $3601181498092$ & $-2016301996664$ & $-172065071792$ & $-3426195952892$ \\
$b_{8a}$ & $-28384454336$ & $201765139520$ & $-285164336240$ & $-798999469184$ & $-9990285620$ & $-256051490576$ \\
$b_{9a}$ & $-194465674912$ & $105361594336$ & $-72681574708$ & $-235689448576$ & $-374891590$ & $114501354956$ \\
$b_{10a}$ & $1028937525$ & $-25723438125$ & $17389581825$ & $20867344575$ & $160123425$ & $15122790375$ \\
$b_{11a}$ & $-337377040$ & $8434426000$ & $-8286998720$ & $-11711539840$ & $-249509260$ & $-8268420160$ \\
$b_{12a}$ & $-337377040$ & $8434426000$ & $-7188621440$ & $-11711539840$ & $-249509260$ & $-8268420160$ \\
$b_{13a}$ & $2206700496$ & $-899026128$ & $0$ & $11881344384$ & $92461824$ & $-6287404032$ \\
$b_{14a}$ & $1005612372765$ & $212189652675$ & $1767367502901$ & $624380198715$ & $5854711590$ & $336447267975$ \\
$b_{15a}$ & $54363904$ & $-1359097600$ & $1777033216$ & $54312692224$ & $726048064$ & $-4220776448$ \\
$b_{16a}$ & $127291559616$ & $808074895680$ & $2126104954272$ & $2027689327872$ & $102637344888$ & $2224662395904$ \\
$b_{17a}$ & $-603490423536$ & $-13109176080$ & $-1879000931808$ & $-110031457536$ & $-42262011792$ & $-380448327168$ \\
$b_{18a}$ & $-380959338672$ & $97971580080$ & $-47428799616$ & $70149908736$ & $5634371016$ & $290932912512$ \\
$b_{19a}$ & $0$ & $0$ & $-1098377280$ & $0$ & $0$ & $0$ \\
\end{tabular}
\end{ruledtabular}
\endgroup
\end{table}

The one-loop amplitudes follow from
\begin{equation}
F_a^{(1)}(r)=\frac{F_a^{(0)}}{4\pi^2}\left[c_a+\frac{25}{3}\ln r\right],
\qquad r=\frac{\mu_R}{m_c}.
\label{eq:finite-amplitude-general}
\end{equation}
where the loop factor is included in $F^{(1)}$.

The physical amplitude and hard contractions are
\begin{align}
\mathcal M&=\frac{e e_c g_s^4}{m_c^5}\sqrt{\rho_H}\sum_{n,a}\left[F_{na}^{(0)}+g_s^2F_{na}^{(1)}\right]L_nB_a,\qquad g_s^2=4\pi\alpha_s,\nonumber\\
h_{nm}^{(0)}&=\sum_{ab}F_{na}^{(0)}G_{ab}F_{mb}^{(0)*},\qquad
h_{nm}^{(1)}=\sum_{ab}\left(F_{na}^{(1)}G_{ab}F_{mb}^{(0)*}+F_{na}^{(0)}G_{ab}F_{mb}^{(1)*}\right).
\label{eq:amplitude-contraction}
\end{align}
In Eq.~\eqref{eq:amplitude-contraction} the operator index $n$ is suppressed in the single-operator axial and tensor channels.

The scalar structure is $B_0=(g^{\mu\nu}-P_H^\mu k^\nu/(P_H\cdot k))\epsilon_{\gamma\mu}^*\epsilon_{H\nu}^*$, and the axial structure is $B_1=-\epsilon_T\cdot\epsilon_\gamma^*+(\epsilon_T\cdot k)(P_T\cdot\epsilon_\gamma^*)/(P_T\cdot k)$. Their metrics are $G_0=2$ and $G_1=2/3$. The tensor has $G_2=\operatorname{diag}(4/5,2/5,2/5)$. These include the initial-spin average. Together with the two-body factor $3/(256\pi m_c)$ and the state conversion, they yield Eq.~\eqref{eq:factorization} of the Letter.

The exact Born coefficients and rounded one-loop checkpoints at $r=\mu_R/m_c=1$ are
\begin{equation}
\begin{array}{c|c|r@{\thickspace}l}
a&F_a^{(0)}&\multicolumn{2}{c}{F_a^{(1)}(1)}\\ \hline
(0,\bar3)&-\frac{4901}{270\sqrt6}&-0.908237603089&-0.037773713892i\\
(0,6)&\frac{95}{108}&-0.399434949411&-0.064840866857i\\
1&\frac{2197\sqrt3}{270}&2.572871388620&+0.062611998493i\\
(2,1)&\frac{454}{135}&0.566544109125&-0.026540852433i\\
(2,2)&\frac{233}{270\sqrt3}&0.087987842468&-0.006098622831i\\
(2,3)&-\frac{46}{27}&-0.478329577277&+0.017513903766i
\end{array}.
\label{eq:finite-numerical}
\end{equation}

\section{Explicit finite coefficients}
\label{sec:explicit-analytic}
Using the elementary functions and dilogarithm combinations defined in Sec.~\ref{sec:definitions}, Eqs.~\eqref{eq:elementary-bases}--\eqref{eq:dilog-combinations}, the six finite coefficients are given below in fully expanded form, without recourse to Tables~\ref{tab:elementary-weights} and~\ref{tab:dilog-weights}. The axial factor $s_1=135/140608$ of Eq.~\eqref{eq:finite-amplitude} is already included in $c_1$.

\subsection{Scalar channel}
\textit{Scalar antitriplet component}. For $F_{(0,\bar3)}^{(0)}=-\frac{4901}{270\sqrt6}$,
\nopagebreak[4]
\begin{align}
c_{(0,\bar3)}={}&\frac{19705408}{4631445}-\frac{57013854379}{93636702432}\pi^2+\frac{403532057}{466849656}\ell_2-\frac{2001013}{743148432}\ell_2\ell_{11}+\frac{3414669857593}{343334575584}\ell_2^2\nonumber\\
&\quad+\frac{105325}{75029409}\ell_2\ell_{23}+\frac{318182959}{389041380}\ell_3-\frac{811954208665}{82873863072}\ell_2\ell_3+\frac{7275333997643}{2403342029088}\ell_3^2-\frac{84023}{411684}\ell_5\nonumber\\
&\quad-\frac{255868253393}{400557004848}\ell_2\ell_5-\frac{586335896801}{1201671014544}\ell_3\ell_5+\frac{8373206987}{100139251212}\ell_5^2-\frac{728363}{743148432}\ell_7\ell_{11}\nonumber\\
&\quad+\frac{45394720867}{92436231888}\ell_2\ell_7+\frac{190216051555}{1201671014544}\ell_3\ell_7-\frac{21065}{75029409}\ell_5\ell_7-\frac{134064519937}{1201671014544}\ell_7^2+\frac{145040L}{44109\sqrt3}\nonumber\\
&\quad-\frac{9432925}{4234464}L^2+\frac{125631767}{111154680}A+\frac{9}{4901}\ell_{11}(\ell_{11}-\ell_3)-\frac{636325}{743148432}U-\frac{21065}{75029409}\ell_{23}(\ell_{23}-\ell_5)\nonumber\\
&\quad+\frac{1771795}{4234464}L(\ell_2+5\ell_3)\nonumber\\
&{}+i\pi\bigl[\frac{128796943}{875343105}+\frac{385}{88218\sqrt3}+\frac{636325}{371574216}\ell_{19}+\frac{2882768459}{6827676219}\ell_2+\frac{17}{1058616}\ell_3-\frac{759412700}{6827676219}\ell_5\nonumber\\
&\qquad-\frac{1077152}{10718487}\ell_7-\frac{880}{132327}L\bigr]\nonumber\\
&\quad-\frac{989731}{6243874}\Psi_{1}-\frac{525723811}{8403293808}\Psi_{2}+\frac{265181}{471613428}\Psi_{3}+\frac{97264719247}{200278502424}\Psi_{4}\nonumber\\
&\quad+\frac{5535310622767}{1201671014544}\Psi_{5}-\frac{464279}{550215666}\Psi_{6}+\frac{418342686661}{75104438409}\Psi_{7}-\frac{253432628}{10729205487}\Psi_{8}\nonumber\\
&\quad-\frac{419107058}{2589808221}\Psi_{9}+\frac{636325}{743148432}\Psi_{10}-\frac{21065}{75029409}\Psi_{11}-\frac{21065}{75029409}\Psi_{12}\nonumber\\
&\quad+\frac{9}{4901}\Psi_{13}+\frac{1771795}{2117232}\Psi_{14}+\frac{485392}{10729205487}\Psi_{15}+\frac{7555292}{71324253}\Psi_{16}\nonumber\\
&\quad-\frac{465191}{926289}\Psi_{17}-\frac{989731}{3121937}\Psi_{18}.
\label{eq:explicit-scalar-antitriplet}
\end{align}

\par\smallskip
\textit{Scalar sextet component}. For $F_{(0,6)}^{(0)}=\frac{95}{108}$,
\nopagebreak[4]
\begin{align}
c_{(0,6)}={}&\frac{17350}{3591}-\frac{768763063}{259290720}\pi^2+\frac{1016561263}{45246600}\ell_2+\frac{16464109}{72025200}\ell_2\ell_{11}+\frac{21494572709}{1751349600}\ell_2^2-\frac{105325}{290871}\ell_2\ell_{23}\nonumber\\
&\quad+\frac{15283949}{1508220}\ell_3+\frac{471411350261}{232929496800}\ell_2\ell_3+\frac{695019422029}{232929496800}\ell_3^2+\frac{5179}{7980}\ell_5-\frac{10675072799}{1552863312}\ell_2\ell_5\nonumber\\
&\quad-\frac{134336682923}{23292949680}\ell_3\ell_5+\frac{195658747}{388215828}\ell_5^2-\frac{15352141}{72025200}\ell_7\ell_{11}-\frac{22204271851}{8958826800}\ell_2\ell_7\nonumber\\
&\quad-\frac{53389496683}{116464748400}\ell_3\ell_7+\frac{21065}{290871}\ell_5\ell_7+\frac{84289501273}{116464748400}\ell_7^2-\frac{308L}{171\sqrt3}-\frac{508903}{82080}L^2-\frac{57680083}{2154600}A\nonumber\\
&\quad-\frac{11}{1425}\ell_{11}(\ell_{11}-\ell_3)+\frac{636325}{2881008}U+\frac{21065}{290871}\ell_{23}(\ell_{23}-\ell_5)+\frac{44863}{49248}L(\ell_2+5\ell_3)\nonumber\\
&{}+i\pi\bigl[-\frac{12636587}{16967475}+\frac{7}{342\sqrt3}-\frac{636325}{1440504}\ell_{19}-\frac{996504107}{661731525}\ell_2-\frac{17}{4104}\ell_3+\frac{154343341}{132346305}\ell_5\nonumber\\
&\qquad+\frac{34144}{207765}\ell_7-\frac{16}{513}L\bigr]\nonumber\\
&\quad+\frac{1374461}{3267810}\Psi_{1}+\frac{327464491}{814438800}\Psi_{2}-\frac{265181}{1828332}\Psi_{3}+\frac{24679098253}{3882158280}\Psi_{4}\nonumber\\
&\quad-\frac{179021864863}{116464748400}\Psi_{5}+\frac{464279}{2133054}\Psi_{6}-\frac{5549133496}{7279046775}\Psi_{7}+\frac{360294892}{207972765}\Psi_{8}\nonumber\\
&\quad+\frac{6585099646}{7279046775}\Psi_{9}-\frac{636325}{2881008}\Psi_{10}+\frac{21065}{290871}\Psi_{11}+\frac{21065}{290871}\Psi_{12}-\frac{11}{1425}\Psi_{13}\nonumber\\
&\quad+\frac{44863}{24624}\Psi_{14}-\frac{485392}{41594553}\Psi_{15}+\frac{9592532}{1382535}\Psi_{16}-\frac{2021}{17955}\Psi_{17}+\frac{1374461}{1633905}\Psi_{18}.
\label{eq:explicit-scalar-sextet}
\end{align}

At $\mu_R=2m_c$, the pure-sextet correction is
$\Gamma_{\rm NLO}/\Gamma_{\rm LO}-1\simeq-1.93$, i.e.
$\Gamma_{\rm NLO}/\Gamma_{\rm LO}\simeq-0.93$. The fixed-order
expansion is therefore not uniformly controlled in this direction.
The widths quoted in the Letter combine both directions and their
interference.

\par\smallskip
\subsection{Axial channel}
\textit{Axial component}. For $F_{1}^{(0)}=\frac{2197\sqrt3}{270}$,
\nopagebreak[4]
\begin{align}
c_{1}={}&\frac{16052327}{4152330}+\frac{7015934041}{167900294016}\pi^2+\frac{419180443}{418554864}\ell_2-\frac{10754225}{1332542016}\ell_2\ell_{11}+\frac{1307398287323}{205211470464}\ell_2^2\nonumber\\
&\quad+\frac{612485}{33633873}\ell_2\ell_{23}+\frac{237545047}{697591440}\ell_3-\frac{21163435187693}{4309440879744}\ell_2\ell_3+\frac{1638184539913}{1436480293248}\ell_3^2-\frac{478853}{1245699}\ell_5\nonumber\\
&\quad-\frac{2276199242171}{2154720439872}\ell_2\ell_5-\frac{459443697563}{718240146624}\ell_3\ell_5+\frac{22378953607}{119706691104}\ell_5^2+\frac{10754225}{1332542016}\ell_7\ell_{11}\nonumber\\
&\quad+\frac{15574666615}{165747726144}\ell_2\ell_7+\frac{715071965}{79804460736}\ell_3\ell_7-\frac{16030}{4804839}\ell_5\ell_7+\frac{50186459311}{718240146624}\ell_7^2+\frac{677375L}{158184\sqrt3}\nonumber\\
&\quad-\frac{18979613}{7592832}L^2+\frac{66021763}{99655920}A-\frac{10754225}{1332542016}U-\frac{129355}{33633873}\ell_{23}(\ell_{23}-\ell_5)+\frac{9341809}{22778496}L(\ell_2+5\ell_3)\nonumber\\
&\quad+\frac{1270}{1245699}\ell_3\ell_{23}\nonumber\\
&{}+i\pi\bigl[-\frac{13540663}{784790370}-\frac{250}{19773\sqrt3}+\frac{10754225}{666271008}\ell_{19}+\frac{803337779}{1360303308}\ell_2+\frac{132229}{4429152}\ell_3-\frac{6012145}{302289624}\ell_5\nonumber\\
&\qquad-\frac{73768}{369603}\ell_7+\frac{500}{59319}L\bigr]\nonumber\\
&\quad-\frac{415867}{37786203}\Psi_{1}+\frac{2685323713}{15067975104}\Psi_{2}+\frac{5648047}{563767776}\Psi_{3}+\frac{224445296147}{359120073312}\Psi_{4}\nonumber\\
&\quad+\frac{1759443607091}{718240146624}\Psi_{5}-\frac{14122861}{845651664}\Psi_{6}+\frac{900295374523}{538680109968}\Psi_{7}-\frac{2546110145}{19238575356}\Psi_{8}\nonumber\\
&\quad-\frac{18170393677}{538680109968}\Psi_{9}+\frac{10754225}{1332542016}\Psi_{10}-\frac{129355}{33633873}\Psi_{11}-\frac{16030}{4804839}\Psi_{12}\nonumber\\
&\quad+\frac{9341809}{11389248}\Psi_{14}+\frac{566656}{687091977}\Psi_{15}+\frac{124451}{126126}\Psi_{16}-\frac{103457}{118638}\Psi_{17}\nonumber\\
&\quad-\frac{831734}{37786203}\Psi_{18}-\frac{635}{1245699}\Psi_{19}.
\label{eq:explicit-axial}
\end{align}

\par\smallskip
\subsection{Tensor channel}
\textit{First tensor component}. For $F_{(2,1)}^{(0)}=\frac{454}{135}$,
\nopagebreak[4]
\begin{align}
c_{(2,1)}={}&\frac{3321163}{1103220}+\frac{360597722209}{624524898816}\pi^2+\frac{151819933}{111204576}\ell_2-\frac{222780197}{4956546816}\ell_2\ell_{11}+\frac{1342472275387}{208174966272}\ell_2^2\nonumber\\
&\quad+\frac{5027275}{62552574}\ell_2\ell_{23}-\frac{11676467}{11583810}\ell_3-\frac{1161536130707}{1457224763904}\ell_2\ell_3-\frac{1508389675015}{1457224763904}\ell_3^2-\frac{11036957}{18534096}\ell_5\nonumber\\
&\quad-\frac{328402317563}{242870793984}\ell_2\ell_5-\frac{494740059371}{728612381952}\ell_3\ell_5+\frac{2080537483}{40478465664}\ell_5^2+\frac{61129253}{4956546816}\ell_7\ell_{11}\nonumber\\
&\quad+\frac{20791021297}{56047106304}\ell_2\ell_7+\frac{87987509569}{728612381952}\ell_3\ell_7-\frac{1005455}{62552574}\ell_5\ell_7+\frac{39701625641}{728612381952}\ell_7^2+\frac{1144415L}{196128\sqrt3}\nonumber\\
&\quad-\frac{32919995}{9414144}L^2-\frac{190510889}{46335240}A+\frac{1799}{110322}\ell_{11}(\ell_{11}-\ell_3)-\frac{141954725}{4956546816}U-\frac{1005455}{62552574}\ell_{23}(\ell_{23}-\ell_5)\nonumber\\
&\quad+\frac{12101095}{28242432}L(\ell_2+5\ell_3)\nonumber\\
&{}+i\pi\bigl[\frac{54522751}{417017160}-\frac{140}{681\sqrt3}+\frac{141954725}{2478273408}\ell_{19}+\frac{3428049208}{2846142117}\ell_2+\frac{21263}{261504}\ell_3-\frac{29791891925}{91076547744}\ell_5\nonumber\\
&\qquad-\frac{30420025}{71488656}\ell_7+\frac{700}{6129}L\bigr]\nonumber\\
&\quad+\frac{375889}{7808346}\Psi_{1}+\frac{24028243093}{56047106304}\Psi_{2}+\frac{1553483}{17872164}\Psi_{3}+\frac{87278032669}{121435396992}\Psi_{4}\nonumber\\
&\quad+\frac{1144961541037}{728612381952}\Psi_{5}-\frac{1644821}{10425429}\Psi_{6}-\frac{252037749583}{91076547744}\Psi_{7}-\frac{127391497}{116169066}\Psi_{8}\nonumber\\
&\quad-\frac{1841323817}{5692284234}\Psi_{9}+\frac{141954725}{4956546816}\Psi_{10}-\frac{1005455}{62552574}\Psi_{11}-\frac{1005455}{62552574}\Psi_{12}\nonumber\\
&\quad+\frac{1799}{110322}\Psi_{13}+\frac{12101095}{14121216}\Psi_{14}+\frac{30308422}{406591731}\Psi_{15}+\frac{22565987}{8108667}\Psi_{16}\nonumber\\
&\quad-\frac{6479}{42903}\Psi_{17}+\frac{375889}{3904173}\Psi_{18}.
\label{eq:explicit-tensor-1}
\end{align}

\par\smallskip
\textit{Second tensor component}. For $F_{(2,2)}^{(0)}=\frac{233}{270\sqrt3}$,
\nopagebreak[4]
\begin{align}
c_{(2,2)}={}&\frac{810182}{660555}+\frac{11694988651}{6677418384}\pi^2-\frac{381162167}{33291972}\ell_2-\frac{1718267}{52995384}\ell_2\ell_{11}-\frac{37321050089}{2225806128}\ell_2^2\nonumber\\
&\quad+\frac{3427325}{21401982}\ell_2\ell_{23}-\frac{92832356}{13871655}\ell_3+\frac{166011001781}{7790321448}\ell_2\ell_3-\frac{161564465269}{15580642896}\ell_3^2+\frac{2668901}{1585332}\ell_5\nonumber\\
&\quad+\frac{1438024657}{2596773816}\ell_2\ell_5-\frac{9906289751}{7790321448}\ell_3\ell_5-\frac{337816123}{288530424}\ell_5^2+\frac{460283}{52995384}\ell_7\ell_{11}-\frac{388832957}{599255496}\ell_2\ell_7\nonumber\\
&\quad-\frac{933732461}{7790321448}\ell_3\ell_7-\frac{685465}{21401982}\ell_5\ell_7+\frac{369054289}{3895160724}\ell_7^2+\frac{122945L}{4194\sqrt3}-\frac{140945}{50328}L^2-\frac{3839771}{7926660}A\nonumber\\
&\quad+\frac{224}{18873}\ell_{11}(\ell_{11}-\ell_3)-\frac{4675}{227448}U-\frac{685465}{21401982}\ell_{23}(\ell_{23}-\ell_5)+\frac{56735}{150984}L(\ell_2+5\ell_3)\nonumber\\
&{}+i\pi\bigl[\frac{60539053}{499379580}-\frac{5320}{2097\sqrt3}+\frac{4675}{113724}\ell_{19}-\frac{1646246209}{973790181}\ell_2+\frac{1199}{12582}\ell_3+\frac{234818800}{973790181}\ell_5\nonumber\\
&\qquad+\frac{351389}{3057426}\ell_7+\frac{7280}{6291}L\bigr]\nonumber\\
&\quad+\frac{966113}{2671578}\Psi_{1}+\frac{128404699}{599255496}\Psi_{2}+\frac{1024763}{12229704}\Psi_{3}+\frac{1663823293}{1298386908}\Psi_{4}\nonumber\\
&\quad-\frac{9605731379}{1947580362}\Psi_{5}-\frac{1934137}{14267988}\Psi_{6}-\frac{21508133974}{973790181}\Psi_{7}-\frac{50970845}{39746538}\Psi_{8}\nonumber\\
&\quad-\frac{187445795}{3895160724}\Psi_{9}+\frac{4675}{227448}\Psi_{10}-\frac{685465}{21401982}\Psi_{11}-\frac{685465}{21401982}\Psi_{12}\nonumber\\
&\quad+\frac{224}{18873}\Psi_{13}+\frac{56735}{75492}\Psi_{14}+\frac{12965144}{139112883}\Psi_{15}+\frac{36551761}{2774331}\Psi_{16}\nonumber\\
&\quad-\frac{238898}{44037}\Psi_{17}+\frac{966113}{1335789}\Psi_{18}.
\label{eq:explicit-tensor-2}
\end{align}

\par\smallskip
\textit{Longitudinal tensor component}. For $F_{(2,3)}^{(0)}=-\frac{46}{27}$,
\nopagebreak[4]
\begin{align}
c_{(2,3)}={}&\frac{84379}{39123}+\frac{63129626233}{45198466560}\pi^2-\frac{215690593}{78871968}\ell_2-\frac{60104669}{2511025920}\ell_2\ell_{11}-\frac{6579777773}{105463088640}\ell_2^2\nonumber\\
&\quad+\frac{1419715}{12675852}\ell_2\ell_{23}-\frac{79368337}{32863320}\ell_3+\frac{5076646284817}{738241620480}\ell_2\ell_3-\frac{3468300386959}{738241620480}\ell_3^2+\frac{942637}{938952}\ell_5\nonumber\\
&\quad+\frac{5061503627}{5349576960}\ell_2\ell_5+\frac{3349154077}{369120810240}\ell_3\ell_5-\frac{152119441}{253169280}\ell_5^2+\frac{145647581}{2511025920}\ell_7\ell_{11}\nonumber\\
&\quad-\frac{5378427043}{5678781696}\ell_2\ell_7-\frac{158781363239}{369120810240}\ell_3\ell_7-\frac{283943}{12675852}\ell_5\ell_7-\frac{6442551725}{73824162048}\ell_7^2+\frac{222817L}{19872\sqrt3}\nonumber\\
&\quad-\frac{2960203}{953856}L^2-\frac{45455099}{9389520}A-\frac{476}{27945}\ell_{11}(\ell_{11}-\ell_3)-\frac{894575}{21835008}U-\frac{283943}{12675852}\ell_{23}(\ell_{23}-\ell_5)\nonumber\\
&\quad+\frac{1304135}{2861568}L(\ell_2+5\ell_3)\nonumber\\
&{}+i\pi\bigl[-\frac{44489323}{73942470}+\frac{7}{207\sqrt3}+\frac{894575}{10917504}\ell_{19}-\frac{26477184683}{23070050640}\ell_2+\frac{12595}{79488}\ell_3-\frac{8983376405}{18456040512}\ell_5\nonumber\\
&\qquad+\frac{7417979}{9054180}\ell_7+\frac{7}{414}L\bigr]\nonumber\\
&\quad+\frac{3117851}{7911540}\Psi_{1}-\frac{12752699843}{28393908480}\Psi_{2}+\frac{3431165}{57946752}\Psi_{3}-\frac{4968290597}{61520135040}\Psi_{4}\nonumber\\
&\quad-\frac{225633278099}{369120810240}\Psi_{5}-\frac{6491689}{67604544}\Psi_{6}-\frac{856548988223}{92280202560}\Psi_{7}-\frac{326596289}{470817360}\Psi_{8}\nonumber\\
&\quad+\frac{28625338739}{92280202560}\Psi_{9}+\frac{894575}{21835008}\Psi_{10}-\frac{283943}{12675852}\Psi_{11}-\frac{283943}{12675852}\Psi_{12}\nonumber\\
&\quad-\frac{476}{27945}\Psi_{13}+\frac{1304135}{1430784}\Psi_{14}-\frac{2355344}{205982595}\Psi_{15}+\frac{24758084}{4107915}\Psi_{16}-\frac{974}{945}\Psi_{17}\nonumber\\
&\quad+\frac{3117851}{3955770}\Psi_{18}.
\label{eq:explicit-tensor-3}
\end{align}

\endgroup

\section{Tensor polarization basis}
\label{sec:polarization}
Choose the photon along $+z$ in the tetraquark rest frame. In units of $m_c$, $k=(3/2,0,0,3/2)$ and $P_\psi=(5/2,0,0,-3/2)$. With $g=\operatorname{diag}(1,-1,-1,-1)$, the vector polarizations are $V_x=(0,1,0,0)$, $V_y=(0,0,1,0)$, and $V_L=(3/4,0,0,-5/4)$. The photon uses $V_x,V_y$. For spatial unit vectors $x,y,z$, the normalized tensor basis is
\begin{equation}
S_1=(xx-yy)/\sqrt2,\quad S_2=(xx+yy-2zz)/\sqrt6,\quad S_3=(xy+yx)/\sqrt2,\quad
S_4=(xz+zx)/\sqrt2,\quad S_5=(yz+zy)/\sqrt2,
\end{equation}
with outer products understood and zero temporal components. The three tensor entries of Eq.~\eqref{eq:finite-numerical} are the amplitudes $(a,b,c)$ for $(S_1,V_x,\gamma_x)$, $(S_2,V_x,\gamma_x)$, and $(S_4,V_L,\gamma_x)$. All nonzero rows are
\begin{equation}
\begin{array}{c|rrrrrrrr}
(S,V,\gamma)&(1,x,x)&(1,y,y)&(2,x,x)&(2,y,y)&(3,x,y)&(3,y,x)&(4,L,x)&(5,L,y)\\ \hline
\mathcal A&a&-a&b&b&a&a&c&c
\end{array}.
\end{equation}
\begin{equation}
\frac15\sum|\mathcal A|^2=\frac45|a|^2+\frac25|b|^2+\frac25|c|^2.
\end{equation}
The longitudinal part is $2|c|^2/5$, giving the metrics in the Letter. At NLO each squared amplitude retains only Born--one-loop interference. The angular and mixed-sample relations then follow from Eqs.~\eqref{eq:polarization-expanded}, \eqref{eq:lepton-angle}, and \eqref{eq:polarization-mixture}.

\section{Numerical prescriptions and production interface}
\label{sec:production-interface}
The coupling follows two-loop running with $\beta_0=25/3$, $\beta_1=154/3$, and $\alpha_s(3~\mathrm{GeV})=0.249$, using fixed $n_f=4$ without threshold switching. Mass scans keep the absolute coupling anchor, LDMEs, and matching ratios fixed while setting $\mu_R=2m_c$. No model-LDME mass dependence or binding corrections are included. Models I and II retain the labels of Ref.~\cite{Feng:2023agq}, based on Refs.~\cite{Lu:2020cns,liu:2020eha}. Spin-summed axial and tensor inputs are divided by three and five. The scalar LDME matrices in Table~\ref{tab:tetraquark-ldmes} are rank one up to rounding. The rounded entries are used directly, without a positive-semidefinite projection. The second tensor LDME is rounded from $0.01256$ to $0.0126~\mathrm{GeV}^9$~\cite{Yang:2026bjy}. The relation $\rho_\eta=\rho_\psi/3$ uses leading spin symmetry with unquantified velocity corrections.

For $M=6.847$ GeV, $m_c=M/4$, $\mu_R=M$, and the spin-summed $\rho_\psi=1.32~\mathrm{GeV}^3$,
\begin{equation}
\alpha_s(M)=0.19131,\qquad H_2^{\rm NLO}=6.125\times10^{-4}~\mathrm{GeV}^{-8},\qquad
\Gamma_{\gamma\psi}=H_2^{\rm NLO}\rho_{T,2}^{\rm fix}.
\end{equation}
For $\rho_{T,2}^{\rm fix}=0.0144~\mathrm{GeV}^9$ this gives $\Gamma_{\gamma\psi}=8.8\times10^{-3}$~MeV, an order of magnitude below the $m_c=1.5$~GeV benchmark width in Table~\ref{tab:central-widths}. The partonic prescription $m_c=M/4$ does not reproduce both physical hadron masses. The production fit retains its published coupling and PDF inputs~\cite{Wang:2025hex}. Sharing $M$ and a central scale does not establish full compatibility. The tensor production current printed in Ref.~\cite{Wang:2025hex} [cf. Eq.~(6) therein] includes the symmetric projection, but the full symmetric traceless projector obeys $\epsilon_{ij}^*\Pi^{ij,kl}=\epsilon^{*kl}$. Its factors of $1/2$ therefore do not imply an overall current rescaling. The paired production coefficient and state convention still require verification. We therefore retain the unverified conversion $\rho_{T,2}^{\rm prod}=5\xi\rho_{T,2}^{\rm fix}$, with $\xi>0$, and count the decay spin average once. The conversion covers operator, state, and polarization conventions. Changing production inputs requires a compatible response and refit. The fit and radiative ratio must describe the same resolved state. Unknown production/decay correlations preclude a complete combined uncertainty.

\FloatBarrier